\documentclass{aa}

\usepackage{natbib}
\usepackage{threeparttable}
\usepackage{graphicx}
\usepackage{subcaption}
\usepackage{txfonts}
\usepackage[]{hyperref}
\newcommand{\G}{\mathcal{G}}

\newcommand{\Ts}{\rm T_\star}

\begin{document} 

  \title{TATOO: Tidal-chronology standalone tool to estimate the age of massive close-in planetary systems}
  \author{F. Gallet\inst{1} 
          }

\offprints{F. Gallet,\\ email: florian.gallet1@univ-grenoble-alpes.fr}

  \institute{$^1$ Univ. Grenoble Alpes, CNRS, IPAG, 38000 Grenoble, France \\
             }

  \date{Received -- ; Accepted --} 

  \abstract
{The presence of a massive close-in planet with an orbital period of a few days or less around a low-mass star can possibly result in a strong variation in the properties of the central star. Indeed, star-planet tidal interactions generate exchanges of angular momentum that can result in tidal spin-up. This effect could then lead to gyrochronological ages biased towards younger ages.} 
{This article provides the community with TATOO, a standalone tool based on tidal-chronology, with which to estimate the age of a massive close-in planetary system using only its observed properties: mass of the planet and the star, stellar rotation, and planetary orbital periods.} 
{I used a star-planet tidal evolution numerical code to create a large multi-parametric grid of the evolution of synthetic star-planet systems. Furthermore, using the tidal-chronology technique, I employed a 3D interpolation method to provide a fairly precise age estimate of any given planetary system composed of one massive close-in planet.}
{About half of the planetary systems investigated in this work are subject to tidal spin-up bias. I pointed out that this bias linearly scales with the ratio between rotation and orbital period, making this quantity a useful proxy to rapidly investigate whether tidal-chronology needs to be used. Moreover, while being model dependent, TATOO can also be used even if no rotational departure is present. In that case, it gives results in agreement with the classical gyrochronological analysis. }
{TATOO is a useful tool specifically designed for massive close-in planetary systems that can also be used as a classical gyrochronological tool. For now it is the {only publicly available software} to estimate the age of massive close-in planetary systems subject to tidal spin-up. In that sense, tidal-chronology can be seen as a first order correction of the impact of tidal interaction on gyrochronology.} 


\maketitle

\section{Introduction}

Together with rotation, radius, and luminosity, age is a fundamental physical parameter for both stellar and exoplanetary fields, and its determination for a given star or planetary system is currently a key issue in astrophysics. In stellar physics, stellar rotation has been found to be involved in most of the physical mechanisms: magnetic field strength via dynamo effect \citep{Jouve07}, orbital evolution in planetary systems \citep{Bolmont16,Gallet18}, and stellar angular momentum evolution \citep{GB13,GB15}. Estimating the age of an isolated star or a star that belongs to a planetary system can thus help to constrain its internal structure and chemical species content (e.g. light elements such as Lithium), through the use of stellar evolution models, and to put boundary conditions on planet formation and migration models \citep{Ida08,Mordasini09,Mordasini12,Alibert13,Amard16,Gallet17,Amard19}. It also has an important role in adding constraints on the star-planet interaction efficiency and characteristic timescale \citep{Lanza11} that could help improve star-planet magnetic and tidal interaction scenarios \citep{Strugarek17,Gallet19,Benbakoura19}.

{However, stellar ages cannot be directly measured, except for the Sun, and as a consequence have to be inferred either from theoretical models (e.g. nucleocosmochronology) or by using empirical laws. {Gyrochronology \citep{Barnes03,Barnes07,Barnes10,Angus19} and magnetochronology \citep{Vidotto14} are two examples of empirical methods that use surface rotation period measurements and magnetic field strength estimation \citep[see][for a list of age estimation methods]{Soderblom10}}.}

Other model-dependant techniques can also be used such as isochrones fitting and main sequence turn-off ages, which both require accurate and precise stellar evolution codes. These techniques suffer from the presence of systematics that are linked to fitting accuracy and uncertain distances \citep{Lebreton14}. Moreover, they can only be used if there were no interactions between the star and its close environment during the main sequence phase. Indeed, in the case of past and already finished interactions, it is difficult to quantify the impact of these external interactions on the surface stellar rotation by only using the observations. Hence, it is hard to estimate the true and apparent age of the star. In that case, incompatible age estimations provided by different techniques could be found to be due to this external origin \citep[e.g.][]{Bonnefoy18}. However, for an on-going interaction, for example if a massive planet ($\rm M_p > 1~M_{jup}$) is seen in the vicinity of the star with a separation less than 0.1 au, these techniques can no longer be used since star-planet tidal interactions have most probably modified the evolution of the surface rotation rate during the system's evolution \citep{Gallet18,Qureshi18}. Consequently, and as predicted in the literature \citep{Maxted15,Penev18} and shown in \citet{Gallet19}, the ages of numerous star-planet systems are currently not known and might appear younger than they are because of tidal spin-up effects. In that case, it is yet possible to use tidal-chronology  to get the age of the system from the measurement of the rotation period of the star and the semi-major axis of its orbiting massive planet \citep[see][for a detailed presentation of the technique]{Gallet19}.

In \citet{Gallet19} we introduced and described the tidal-chronology technique. By applying it to the specific system WASP-43, we pointed out the significant discrepancy between tidal-chronological and gyrochronological age estimation induced by tidal spin-up. In \citet{Gallet19} we used a tailored model grid matching the specific observed properties of WASP-43 to perform this estimation. Since this process is time consuming because of the large initial parameter space of planetary systems to be explored, I decided to provide the community with a standalone tool dedicated to the estimation of the age of planetary systems (composed of one low-mass star and one massive close-in planet). This tool uses only the observation of the surface rotation rate of the host star and the current location of the planet orbiting it. This tool is based on a pre-computed generic grid in which the age of the user's system is interpolated, thus making the age estimation process easy and fast.

{This paper is separated as follows. In Sect. \ref{tatoo_principle}, I recall the basic principle of the tidal-chronology technique and describe the standalone tool TATOO  (Tidal-chronology Age TOOl). I then apply this tool to known planetary systems to estimate their age and compare them to classical gyrochronology ones. Finally, I discuss the results in Section \ref{discussion} and conclude in Section \ref{conclusion}.}

\section{TATOO: A standalone tool}
\label{tatoo_principle}

To properly estimate the age of a given system with the tidal-chronology technique (see next section), one needs to construct a dense tailored grid of a planetary system's evolution. This tailored grid should contain the evolution of the stellar rotation period $\rm P_{rot,\star}$ and planetary orbital period $\rm P_{orb,p}$ of a given system that matches the observed stellar $\rm M_\star$ and planetary $\rm M_{p}$ mass, and this for a large range of initial conditions of stellar rotation $\rm P_{\rm rot, init}$ and planetary orbital period $\rm P_{orb,init}$. Hence, for the fixed observed properties $\left\{ \rm M_\star;\rm M_{p} \right\}$ of the system, the tailored grid should explore every possible value of $\rm P_{orb,init}$ for each value of $\rm P_{\rm rot, init}$. {For example, in the case of WASP-43 that we investigated in \citet{Gallet19}, we used a 0.71 $\rm M_{\odot}$ stellar model and a 2.052 $\rm M_{\rm jup}$ mass planet for which we explored the initial conditions of the stellar rotation rate and orbital period. This tailored gird is thus only valid for planetary systems that are identical to the WASP-43 ones.}

The age of the system is then extracted from this tailored grid by using the observed pair $\left\{ \rm P_{rot,obs};\rm P_{orb,obs} \right\}$. However, this procedure is extremely time consuming and the resulting tailored grid is only valid for the considered planetary system corresponding to the observed couple $\left\{ \rm M_\star;\rm M_{p} \right\}$. A generic numerical tool is thus required to provide us with a faster estimate of the age of a given observed system. In this section, I present such a tool named TATOO (https://github.com/GalletFlorian/TATOO) and based on the work of \citet{Gallet19}.

To allow a clear reading of the paper, the nomenclature used in this work is explained below:
\begin{itemize}
        \item tailored grid, initial grid, pre-compiled age exploration grid: These are three different grids that will be introduced in the text.
        \item $\rm P_{rot,\star}$ and $\rm P_{orb,p}$: Without any specific label, they simply refer to the evolution of the stellar rotation and planetary orbital period.
        \item $\rm P_{rot,init}$ and $\rm P_{orb,init}$: The initial stellar rotation period and planetary orbital period used in the dense initial grid. They are distributed between 1-11 days and 0.1-1.1 $\rm P_{rot,init}$, respectively. 
        \item $\rm P_{rot,i}$ and $\rm P_{orb,i}$: The stellar rotation period and planetary orbital period of the exploration grid. They are regularly distributed between 2 and 30 days, and 0.2 and 15 days, respectively (see Sect. \ref{principle}),
        \item $\rm P_{rot,obs}$ and $\rm P_{orb,obs}$: The stellar rotation period and planetary orbital period of the user's favourite observed planetary system.
\end{itemize}
In the next section, I reiterate the general principles of the tidal-chronology technique and how age estimation can be performed using such a technique. 

\subsection{Basic principle of tidal-chronology}

In \citet{Gallet19}, and following the work of \citet{Gallet18}, we presented the concept of a new age estimation technique based only on the measurement of the surface rotation rate of the star and the location of a massive close-in planet around it: the tidal-chronology. In \citet{Gallet19} we demonstrated that the observed pair composed of the stellar rotation period ($\rm P_{rot,obs}$) and planetary orbital period ($\rm P_{orb,obs}$) of a given star and massive close-in planet system is only retrieved at a unique age or during a short range of time. We hence highlighted that this characteristic could be used to estimate the age of massive close-in planetary systems.

As mentioned above, a dense tailored grid of the evolution planetary systems first needs to be created. From this tailored grid, only systems in which the planet is still orbiting the star (i.e. systems that did not experience planetary engulfment) are selected and the ages at which the observed pair $\left\{ \rm P_{\rm rot,obs};\rm P_{orb,obs} \right\}$ is retrieved are extracted. To estimate the validity of this age estimate, in \citet{Gallet19} we introduced the $\rm S^2$ quantity that needs to be minimized
\begin{eqnarray}
\rm S^2 = \frac{(\rm P_{orb,p}-\rm P_{\rm orb,obs})^2}{ \rm \sigma_{\rm P_{orb,obs}}^2} + \frac{(\rm P_{\rm rot,\star}-\rm P_{\rm rot,obs})^2}{ \rm \sigma_{\rm P_{rot,obs}}^2},
\end{eqnarray}
where  $\displaystyle \sigma_{\rm P_{orb,obs}}$ and $\displaystyle \sigma_{\rm P_{rot,obs}}$ are the errors of the observed $\rm  P_{\rm orb,obs}$ and $\rm P_{rot,obs}$, and $\rm P_{rot,\star}$ and $\rm P_{orb,p}$ are the stellar rotation and planetary orbital period from the models. This method finally provides the most probable range for the age of the considered planetary system. 

The model used in this work to create the tailored grid is PROBE (PeRiod and OrBital Evolution code), which is the combination of the stellar angular momentum evolution code JEVOL described in \citet{GB13,GB15} with the modified orbital evolution model used in \citet[][]{Bolmont16}. A detailed description of this model can be found in \citet[][]{Gallet18}. {To enable an easier reading of the paper, I will recall the main characteristics of the model.}

\subsubsection{Tidal dissipation }

{The tidal theory used in this work includes both equilibrium and dynamical tides and it is based on the tidal dissipation formalism described in \citet{Bolmont16} that is parametrized following the simplified model by \citet{Ogilvie13}. As in \citet{Bolmont16} and \citet{Gallet18}, only the frequency-averaged tidal dissipation is considered and the frequency response of the tides is neglected.} {Moreover, the dissipation of tidal inertial waves is only treated inside the convective envelope of the star \citep{Mathis15,Mathis16,Gallet17}}. {For now, the dissipation inside the planet's interior is not included as it is still hardly theoretically constrained. However, for Hot Jupiter cases with circular orbit, we can neglect this additional dissipation regarding the short characteristic timescales ($10^5$ yr) associated to the evolution of both their rotation rate and inclination angle \citep{Leconte10,Damiani18}. Finally, I neglected the impact of the magnetic star-planet interactions on the planetary orbit \citep[see][]{Strugarek17}, while it may have an effect on the rotational evolution of the central star during the early pre-main-sequence (hereafter PMS) phase \citep[][]{Strugarek19}.}
{The migration timescale of planets during the main-sequence (hereafter MS) phase could be shortened by the inclusion of these additional dissipations, which could slightly change the age estimation given by tidal-chronology.}

The impact of the evolution of the planetary orbit on the stellar angular momentum is evaluated using \citet[][]{Gallet17}, who provide an estimation of tidal dissipation effects induced by gravitational interactions between the central star and the orbiting body. {I note that this tidal formalism only allows the presence of one planet orbiting the star with a coplanar and circular orbit.}
 
\subsubsection{Stellar rotation}

{As pointed out in the previous section, the evolution of the planetary semi-major axis $a$ strongly depends on the evolution of the surface rotation rate of the star via its impact on the tidal dissipation intensity \citep{Mathis15,Mathis16,Gallet17,Bolmont17}. As a consequence, in that framework, modelling of the evolution of the stellar rotation should play a crucial role. In PROBE, the rotation rate of the star is modelled using the formalism described in \citet{GB13,GB15}.}

\subsubsection{Stellar model}

{The evolution of the internal structure and the main physical properties of the star are provided by the stellar evolution code STAREVOL \citep[see][and references therein]{Lagardeetal12,Amard16,Amard19}. {A detailed comparison between STAREVOL and similar stellar evolution codes \citep[including MESA, YREC, and PARSEC, see][ respectively]{Choi16,Spada11,Tognelli11} can be found in \citet{Amard19}. However, in this work I did not explore the possible impact of the choice of stellar model on the outcome of this standalone tool. } }

Finally, it is worth noting that as in gyrochronology, the tidal-chronology technique will provide degenerated solutions for a rotational period below ten days, which corresponds to a system younger than about 100 Myr. 

\begin{table}[!h]
\caption{Comparison between $\rm Age_{gyro}$ and $\rm Age_{tidal-gyro}$.} 
\label{tabsystems_gyro}
\centering
        \begin{tabular}{|c|c|c|c|}
        \hline 
         Name & $\rm Age_{gyro}$ & $\rm Age_{tidal-gyro}$ & Deviation \\
         & (Myr) & (Myr) & \%  \\
        \hline 
        \hline
WASP-23 & 1265$\pm$198 & 1269$\pm$176 & 0.3  \\ 
Kepler-423 & 2242$\pm$12 & 2267$\pm$14 & 1.1  \\ 
HD189733 & 695$\pm$1 & 704$\pm$5 & 1.4  \\ 
Qatar-1 & 2516$\pm$54 & 2559$\pm$54 & 1.7  \\ 
CoRoT-13 & 1661$\pm$631 & 1722$\pm$904 & 3.6  \\ 
WASP-46 & 1224$\pm$73 & 1273$\pm$73 & 3.9  \\ 
WASP-124 & 2378$\pm$698 & 2478$\pm$1047 & 4.1  \\ 
WASP-4 & 2498$\pm$57 & 2613$\pm$79 & 4.5  \\ 
HATS-15 & 699$\pm$94 & 733$\pm$165 & 4.8  \\ 
HATS-18 & 771$\pm$43 & 811$\pm$108 & 5.1  \\ 
WASP-57 & 1011$\pm$344 & 950$\pm$381 & 6.2  \\ 
WASP-140 & 634$\pm$32 & 677$\pm$64 & 6.5  \\ 
WASP-36 & 2107$\pm$733 & 2300$\pm$1058 & 8.7  \\ 
WASP-50 & 1416$\pm$51 & 1546$\pm$42 & 8.8  \\ 
NGTS-10 & 1068$\pm$25 & 964$\pm$27 & 10.3  \\ 
WASP-43 & 769$\pm$28 & 854$\pm$30 & 10.4  \\ 
CoRoT-29 & 1160$\pm$249 & 1300$\pm$264 & 11.4  \\ 
TrES-5 & 794$\pm$75 & 892$\pm$92 & 11.7  \\ 
HATS-2 & 841$\pm$39 & 949$\pm$42 & 12.0  \\ 
WASP-19 & 816$\pm$36 & 922$\pm$34 & 12.3  \\ 
HATS-30 & 1627$\pm$232 & 1861$\pm$304 & 13.5  \\ 
HAT-P-37 & 1298$\pm$232 & 1488$\pm$291 & 13.6  \\ 
WASP-10 & 608$\pm$3 & 519$\pm$6 & 15.7  \\ 
HATS-9 & 2094$\pm$395 & 2463$\pm$680 & 16.2  \\ 
HAT-P-53 & 2123$\pm$255 & 2500$\pm$419 & 16.3  \\ 
WASP-41 & 1940$\pm$56 & 2292$\pm$74 & 16.7  \\ 
WASP-80 & 1601$\pm$202 & 1352$\pm$222 & 16.9  \\ 
WASP-65 & 1176$\pm$203 & 1408$\pm$213 & 18.0  \\ 
WASP-135 & 823$\pm$154 & 990$\pm$243 & 18.4  \\ 
WASP-77-A & 1704$\pm$57 & 2130$\pm$72 & 22.2  \\ 
WASP-5 & 1803$\pm$267 & 2265$\pm$379 & 22.7  \\ 
HATS-34 & 956$\pm$158 & 1201$\pm$194 & 22.8  \\ 
WASP-44 & 1363$\pm$408 & 1721$\pm$496 & 23.2  \\ 
HAT-P-36 & 1781$\pm$57 & 2263$\pm$80 & 23.8  \\ 
WASP-85 & 1647$\pm$172 & 2107$\pm$226 & 24.5  \\ 
HATS-14 & 1023$\pm$328 & 1329$\pm$492 & 26.1  \\ 
HATS-33 & 2913$\pm$171 & 3970$\pm$306 & 30.7  \\ 
Qatar-2 & 538$\pm$24 & 390$\pm$47 & 31.9  \\ 
TrES-2 & 3364$\pm$3115 & 2348$\pm$2177 & 35.6  \\ 
HAT-P-43 & 4082$\pm$942 & 6040$\pm$1502 & 38.7  \\ 
WASP-64 & 1648$\pm$452 & 2442$\pm$612 & 38.9  \\
        \hline
        \end{tabular}   
\end{table}

 \begin{figure*}[!ht]
    \centering
    \begin{subfigure}[b]{0.5\linewidth}
        \centering
        \includegraphics[width=\linewidth]{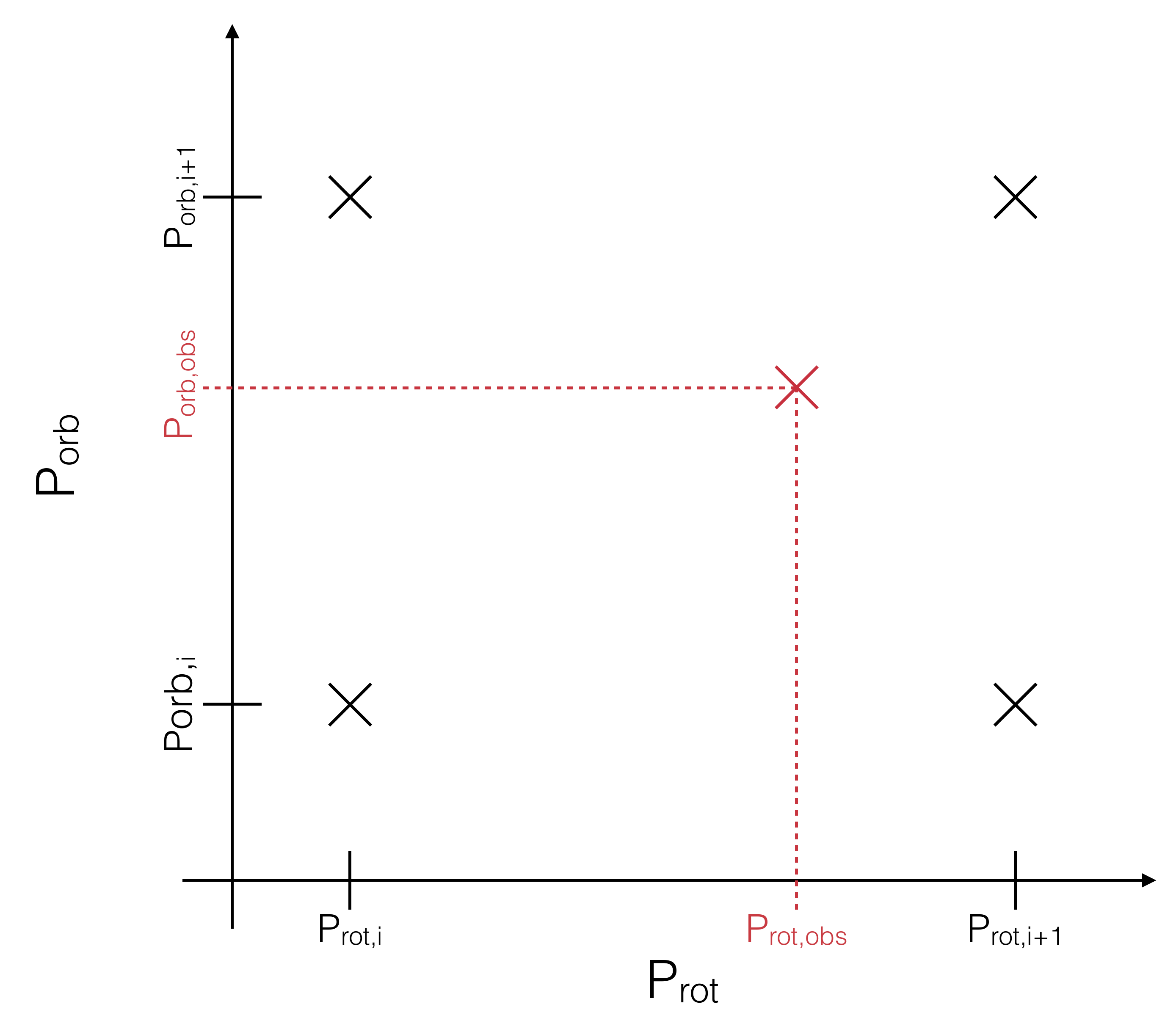}
        \caption{Four-nearest selection.}
    \end{subfigure}%
    ~ 
    \begin{subfigure}[b]{0.5\linewidth}
        \centering
        \includegraphics[width=\linewidth]{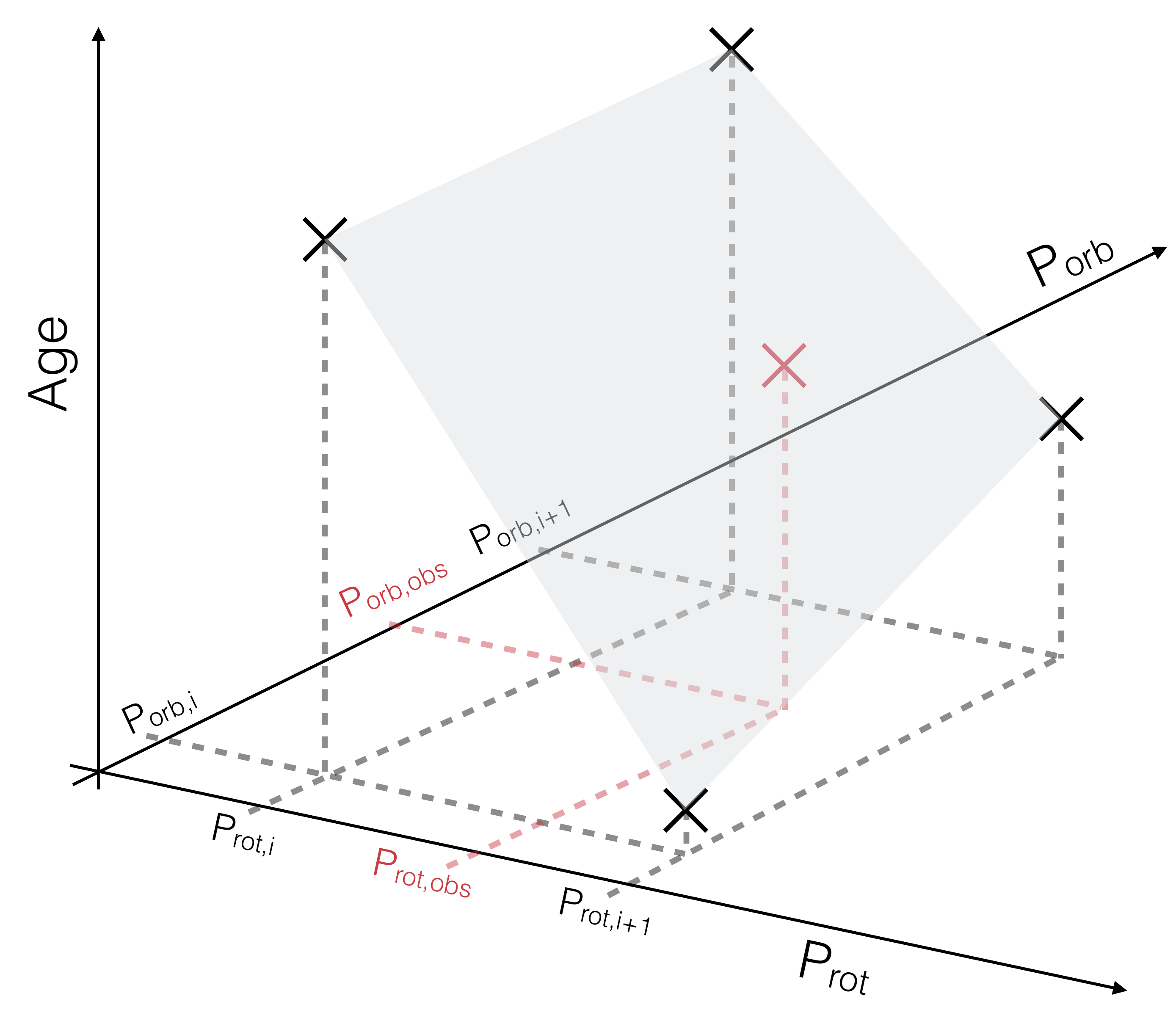}
        \caption{Creation of the three-dimensional map.}
    \end{subfigure}
    \caption{Schematic view of the neighbours technique.}
     \label{schema_all}
\end{figure*}

\subsection{Principle}
\label{principle}
The initial step to use TATOO is first to model a very large range of generic planetary system evolutions (that will be called the initial grid in the following); then, from this initial grid, pre-compiled age exploration files are produced at specific pairs $\left\{  \rm P_{rot,i};\rm P_{orb,i} \right\}$ for each set of $\left\{ \rm M_{\star};\rm M_p \right\}$ masses explored in this work (that will be called the pre-compiled age exploration grid in the following). These specific pairs are regularly spaced in the  $\left\{ \rm P_{rot,\star};\rm P_{orb,p} \right\}$ space. TATOO heavily relies on this pre-compiled exploration grid as it will be used to interpolate the age of a given planetary system by using its observed properties $\left\{ \rm P_{rot,obs};\rm P_{orb,obs} \right\}$. I consider TATOO and PROBE as two different ``entities'' since TATOO can be used with any other grids as long as they follow the format of the current one.

{In contrast to the tailored grid introduced above, the initial grid is designed to be as generic as possible.} It is composed of stars between 0.5 and 1.0 $\rm M_{\odot}$ (with $\rm P_{rot,init}$ between 1 and 11 days) and planetary mass from 0.5 to 3.5 $\rm M_{jup}$ (with $\rm P_{orb,init}$ between 0.1 and 1.0 $\rm P_{rot,init}$). It is produced using the model presented in \citet{Gallet18}.
The range of stellar mass is adopted because the rotational evolution of these stars is well modelled. The initial rotation period $\rm P_{rot,init}$ is chosen to cover the observed range of rotation in the early-PMS cluster (million-year-old clusters  e.g. the Orion Nebulae Cluster and NGC6530). Finally the range of planetary mass between 0.5 to 3.5 is selected so as to ensure a good ratio between a strong enough tidal interaction and sufficiently numerous observed planetary systems in the literature. The upper planetary mass limit can be extended up to the brown dwarf limit (i.e. \textasciitilde\ 13 $\rm M_{jup}$) and the stellar mass range down to 0.3 and up to 1.2 $\rm M_{\odot}$. However, the lower limit of 0.5 $\rm M_{jup}$ is already a strong constraint. Indeed, a massive enough planet is needed in order to have a measurable impact on the rotation of the star through tidal interaction. Section \ref{limits} describes in more detail the limitations of the present tool (Fig. \ref{initialgrid} summarizes how the initial grid is created). {This grid cannot be directly used to extract the age of a given planetary system, except if the properties of this observed system already match one of the synthetic systems from the initial gird. In any other cases, an interpolation first needs to the realized.} 

From this initial grid, I then extract the ages of the systems that fulfil the requirement $\rm S^2 \leq 100$ (voluntarily larger than the actual maximum critical $\rm S^2$ value, see Eq. \ref{ellipse}). This extraction is done for each stellar mass at the specific pairs $\left\{ \rm P_{rot,i};\rm P_{orb,i}\right\}$ and the results are stored in pre-compiled age exploration files as a function of the planet's mass: the pre-compiled age exploration grid.

{It is important to understand here that the initial grid and the pre-compiled age exploration grid are two different grids. The initial grid is composed of the temporal evolution of planetary systems, with given initial conditions, which are produced with the PROBE code. The pre-compiled age exploration grid is composed of the ages found from the initial grid at the specific pairs $\left\{ \rm P_{rot,i};\rm P_{orb,i} \right\}$ for a given $\rm M_{\star}$ star, but for the whole range of planetary masses $\rm M_p$. The interpolation in planetary mass to extract the age of the actual observed system is done in the pre-compiled age exploration grid.  

The relation between the orbital period and the semi-major axis is given by 
\begin{eqnarray}
\rm P_{orb,p} = 2\pi \sqrt{\frac{\rm a^3}{\rm \mathcal{G}(M_{\star}+M_p)}},
\end{eqnarray}
but I note that even if the planetary distances are expressed in $\rm P_{orb,p}$ units, the model itself works with the semi-major axis, which is independent of the properties of the system. Indeed a given value of $\rm P_{orb,p}$ will correspond to different physical distances ($\rm a$) depending on the masses of the star and planet of the system.
In this work $\rm P_{rot,i}$ ranges from 2 to 30 days and $\rm P_{orb,i}$ from 0.36-0.51 days (semi-major axis = 0.01 au) to 10.67-15.11 days (semi-major axis = 0.095 au) are considered. I explored this range of $\rm P_{rot,i}$ for the exploration grid because most of the stars have a surface rotation period within 2-30 days between the PMS and the end of the MS phase \citep[see][]{GB13,GB15}. For $\rm P_{orb,i}$, I investigated planets with a semi-major axis $a$ from 0.095 au down to 0.01 au because below 0.01 au the planet is most probably engulfed by its host star, and beyond 0.05 au no tidal interaction and migration are expected given the planetary masses considered here.

Finally, TATOO follows these steps: \footnote{A detailed tutorial on how to use the tool can be found on the github repository of TATOO https://github.com/GalletFlorian/TATOO/.}
\begin{enumerate} %
        \item It attains the observed periods and masses ($\left\{ \rm P_{rot,obs};\rm P_{orb,obs};\rm M_{p};\rm M_{\star} \right\}$) together with error bars ($\rm \sigma_{P_{rot,obs}}$ and $\rm \sigma_{P_{orb,obs}}$) of the user's planetary system. {A draw is then done on $\rm P_{rot,obs}$ and $\rm P_{orb,obs}$ given $\rm \sigma_{P_{rot,obs}}$ and $\rm \sigma_{P_{orb,obs}}$. These random periods are used as the input $\rm P_{rot,obs}$ and $\rm P_{orb,obs}$.} 
        \item Then TATOO finds the four synthetic pre-compiled age exploration files (defined by the values of $\rm P_{rot,i}$ and $\rm P_{orb,i}$) that encompass the observed $\left\{ \rm P_{rot,obs};\rm P_{orb,obs} \right\}$ pair.
        \item For each of these couples, linear interpolation on the age as a function of planetary mass is performed, which ends up with four numerical relations of the form: $\rm Age =  \theta_1\times M_{p} + \theta_0$ (with $\theta_1$ and $\theta_0$ two numerical constants) and four evaluations of the linearity via the Pearson coefficient $\rm Pe_{c}$. 
        \item TATOO then finds the age of these four couples at the observed planetary mass, which creates a 3D map composed of $\left\{ \rm age;\rm P_{rot,i};\rm P_{orb,i} \right\}$ for the four nearest points. To assess the robustness of the final age estimation process, I introduced a quality number into the rest of this paper that depends on the value $\rm Pe_c$.
        \item Finally, TATOO uses a 3D interpolation method (Python-SciPy griddata routine) to provide the age of the requested system with the observed $\left\{ \rm P_{rot,obs};\rm P_{orb,obs};\rm M_{p};\rm M_{\star} \right\}$ quadruplet.
        \item Procedures 1 to 5 are repeated 100 times for the two stellar masses that border the observed stellar mass $\rm M_{\star}$ and the final age is given using a linear interpolation between the two explored masses. The error on this final age estimate is given from the standard deviation of these 100 age evaluations.
\end{enumerate}
Figure \ref{schema_all} describes the four nearest points selection method. In this framework, the correct definition of the $\rm S^2$ quantity is finally
\begin{eqnarray}
\rm S^2 = \frac{(\rm P_{orb,p}-\rm P_{orb,i})^2}{ \rm \sigma_{\rm P_{orb,p}}^2} + \frac{(\rm P_{\rm rot,\star}-\rm P_{\rm rot,i})^2}{ \rm \sigma_{\rm P_{rot,\star}}^2},
\end{eqnarray}
where $\displaystyle \sigma_{\rm P_{orb,p}} \sim 10^{-5} $ days (error on $a$: $\displaystyle \sigma_{ a}$ = 0.00018 au) and $\displaystyle \sigma_{\rm P_{rot,\star}} = 0.4$ days in this instance. 

Given that the expression of $\rm S^2$ above is expressed in ``standard deviation units'', we can directly use it to define a critical $\rm S^2$ value. Since $\rm S^2$ is here a sum, we can consider an ``ellipse''  of semi-major axis $\rm 3\sigma$  defined by
\begin{eqnarray}
\label{ellipse}
\rm 1 = \left( \frac{  \rm P_{orb,p}-\rm P_{orb,i}   }{ \rm 3\sigma_{\rm P_{orb,p} }} \right)^2 + \left( \frac{  \rm P_{\rm rot,\star}-\rm P_{\rm rot,i}  }{ \rm 3\sigma_{\rm P_{rot,\star}}} \right)^2,
\end{eqnarray}
which then produces a threshold $\rm S_{lim}^2 =9$. In Table \ref{s2_lim} I explored the impact of the choice of $\rm S_{lim}^2$ on the age estimation.

\begin{table*}
\caption{Ages of known massive close-in planetary systems.}
\label{tabsystems}
\centering
        \begin{tabular}{|c|c|c|c|c|c|c|c|c|c|}
        \hline 
        Name & $\rm M_{\star}$ & $\rm M_{p}$ & $\rm P_{rot,obs}$  & $\rm P_{orb,obs}$ & $\rm Age_{gyro}$ & $\rm Age_{tidal}$ & Deviation & $\rm Pe_{c}$ & Quality\\ 
                                &       ($\rm M_{\odot}$) & ($\rm M_{jup}$) & (days)  & (days) & (Myr) & (Myr) & \% & &  \\ 
        \hline 
        \multicolumn{10}{c|}{No tidal spin-up } \\
        \hline
WASP-140   & 0.90 & 2.44 & 10.4$\pm$0.5 & 2.24 & 634$\pm$32 & 564$\pm$124 & -11.8 & 0.33 & 34.55 \\
HATS-15    & 0.87 & 2.17 & 11.1$\pm$1.4 & 1.75 & 699$\pm$94 & 709$\pm$141 & 1.5 & 0.26 & 35.03 \\ 
WASP-80    & 0.58 & 0.54 & 23.5$\pm$3 & 3.07 & 1601$\pm$202 & 1341$\pm$187 & -17.7 & 0.41 & 37.17 \\ 
Qatar-2    & 0.74 & 2.49 & 11.4$\pm$0.5 & 1.34 & 538$\pm$24 & 413$\pm$41 & -26.4 & 0.41 & 40.07 \\ 
WASP-57    & 0.89 & 0.64 & 12.7$\pm$4.5 & 2.84 & 1011$\pm$344 & 957$\pm$518 & -5.6 & 0.51 & 61.85 \\ 
CoRoT-13 & 1.08 & 1.308 & 13.0$\pm$5 & 4.04 & 1661$\pm$631 & 2014$\pm$1009 & 19.2 & 0.76 & 62.00 \\
TrES-5     & 0.90 & 1.79 & 11.6$\pm$1.1 & 1.48 & 794$\pm$75 & 872$\pm$247 & 9.4 & 0.51 & 63.53 \\ 
HATS-34    & 0.95 & 0.94 & 12.2$\pm$1.8 & 2.11 & 956$\pm$158 & 1066$\pm$167 & 10.9 & 0.51 & 66.38 \\ 
WASP-23    & 0.78 & 0.88 & 17.7$\pm$2.7 & 2.94 & 1265$\pm$198 & 1238$\pm$179 & -2.1 & 0.51 & 66.92 \\ 
{HD189733} & 0.82 & 1.142 & 12.0$\pm$0.01 & 2.22 & 695$\pm$1 & 663$\pm$17 & -4.6 & 0.37 & 69.00 \\ 
WASP-4     & 0.89 & 1.22 & 22.2$\pm$0.5 & 1.34 & 2498$\pm$57 & 2695$\pm$164 & 7.6 & 0.64 & 76.43 \\ 
WASP-41    & 0.93 & 0.94 & 18.4$\pm$0.5 & 3.05 & 1940$\pm$56 & 2289$\pm$77 & 16.5 & 0.64 & 89.73 \\ 
TrES-2     & 0.98 & 1.2 & 25.3$\pm$19 & 2.47 & 3364$\pm$3115 & 2354$\pm$2698 & -35.3 & 0.8 & 90.87 \\ 
WASP-124   & 1.07 & 0.6 & 16.1$\pm$4.6 & 3.37 & 2378$\pm$698 & 2643$\pm$1077 & 10.6 & 0.8 & 92.45 \\ 
WASP-10 & 0.76 & 3.16 & 11.9$\pm$0.05 & 3.08 & 608$\pm$3 & 419$\pm$13 & -36.7 & 0.65 & 92.23 \\ 
WASP-50    & 0.89 & 1.47 & 16.3$\pm$0.5 & 1.96 & 1416$\pm$51 & 1662$\pm$50 & 16.0 & 0.64 & 93.24 \\ 
HAT-P-37   & 0.93 & 1.17 & 14.5$\pm$2.6 & 2.80 & 1298$\pm$232 & 1485$\pm$301 & 13.4 & 0.8 & 94.93 \\ 
Kepler-423 & 0.85 & 0.6 & 22.0$\pm$0.12 & 2.68 & 2242$\pm$12 & 2318$\pm$14 & 3.3 & 0.8 & 255.57 \\ 

        \hline 
        \multicolumn{10}{c|}{Tidal spin-up } \\
        \hline
WASP-19    & 0.90 & 1.07 & 11.8$\pm$0.5 & 0.79 & 816$\pm$36 & 3789$\pm$404 & 129.2 & 0.41 & 39.38 \\
WASP-135   & 0.98 & 1.9 & 10.4$\pm$2 & 1.40 & 823$\pm$154 & 1811$\pm$554 & 75.0 & 0.51 & 63.27 \\
CoRoT-29   & 0.97 & 0.85 & 13.0$\pm$2.5 & 2.85 & 1160$\pm$249 & 1453$\pm$292 & 22.5 & 0.64 & 64.98 \\ 
WASP-65    & 0.93 & 1.55 & 14.2$\pm$2.1 & 2.31 & 1176$\pm$203 & 1563$\pm$285 & 28.3 & 0.64 & 65.48 \\ 
HAT-P-53   & 1.09 & 1.48 & 14.9$\pm$2 & 1.96 & 2123$\pm$255 & 3886$\pm$689 & 58.7 & 0.64 & 65.64 \\ 
HATS-30    & 1.09 & 0.71 & 13.0$\pm$1.7 & 3.17 & 1627$\pm$232 & 2027$\pm$293 & 21.9 & 0.51 & 66.92 \\ 
WASP-85 & 1.02 & 1.265 & 14.6$\pm$1.47 & 2.66 & 1647$\pm$172 & 2063$\pm$231 & 22.4 & 0.76 & 68.93 \\ 
WASP-36    & 1.08 & 2.36 & 15.0$\pm$5.5 & 1.54 & 2107$\pm$733 & 5710$\pm$4125 & 92.2 & 0.8 & 91.38 \\ 
HATS-14    & 0.97 & 1.07 & 12.4$\pm$3.9 & 2.77 & 1023$\pm$328 & 1296$\pm$598 & 23.5 & 0.8 & 92.17 \\ 
WASP-64    & 1.00 & 1.27 & 15.8$\pm$3.7 & 1.57 & 1648$\pm$452 & 3554$\pm$1629 & 73.3 & 0.8 & 92.18 \\ 
WASP-43 & 0.71 & 2.052 & 15.6$\pm$0.4 & 0.84 & 769$\pm$28 & 3199$\pm$1461 & 122.5 & 0.8 & 92.19 \\ 
WASP-5     & 0.96 & 1.58 & 17.1$\pm$2.5 & 1.63 & 1803$\pm$267 & 3726$\pm$1264 & 69.5 & 0.8 & 92.95 \\ 
HATS-9     & 1.03 & 0.84 & 16.6$\pm$3.3 & 1.92 & 2094$\pm$395 & 3544$\pm$1069 & 51.5 & 0.8 & 93.32 \\
WASP-44    & 0.95 & 0.89 & 14.7$\pm$4.3 & 2.42 & 1363$\pm$408 & 1692$\pm$469 & 21.5 & 0.8 & 93.61 \\  
HAT-P-43   & 1.05 & 0.66 & 23.2$\pm$4.9 & 3.33 & 4082$\pm$942 & 5967$\pm$1345 & 37.5 & 0.8 & 94.44 \\ 
HATS-18    & 1.04 & 1.98 & 9.4$\pm$0.5 & 0.84 & 771$\pm$43 & 3393$\pm$94 & 126.0 & 0.51 & 96.10 \\ 
HATS-2     & 0.88 & 1.34 & 12.5$\pm$0.5 & 1.35 & 841$\pm$39 & 1357$\pm$35 & 47.0 & 0.51 & 98.77 \\ 
HATS-33    & 1.06 & 1.19 & 19.0$\pm$1 & 2.55 & 2913$\pm$171 & 4260$\pm$328 & 37.6 & 0.8 & 102.99 \\ 
WASP-46    & 0.83 & 1.91 & 16.1$\pm$1 & 1.43 & 1224$\pm$73 & 2153$\pm$127 & 55.0 & 0.8 & 106.95 \\ 
WASP-77-A  & 1.00 & 1.76 & 15.4$\pm$0.5 & 1.36 & 1704$\pm$57 & 4309$\pm$204 & 86.6 & 0.8 & 111.12 \\ 
HAT-P-36   & 1.02 & 1.83 & 15.3$\pm$0.5 & 1.33 & 1781$\pm$57 & 4888$\pm$213 & 93.2 & 0.8 & 112.95 \\ 
Qatar-1    & 0.84 & 1.29 & 23.7$\pm$0.5 & 1.42 & 2516$\pm$54 & 3377$\pm$36 & 29.2 & 0.41 & 123.81 \\ 
NGTS-10 & 0.70 & 2.162 & 17.3$\pm$0.4 & 0.77 & 1068$\pm$25 & 6117$\pm$6 & 140.5 & 0.9 & 1109.50 \\

        \hline
        \end{tabular}
    \begin{tablenotes}
      \small
      \item \textbf{Notes.}     All of these systems are from \citet{Maxted15} and \citet{Penev18}, except NGTS-10 that is from \citet{NGTS10}.
    \end{tablenotes}    
\end{table*}

\subsection{Limitation}
\label{limits}

{In TATOO the age of the system is interpolated from the planetary mass after the interpolations in stellar mass, stellar rotation period, and planetary orbital period. The main problem with this tool is the assumption that there is a linear correlation between the age of the system and the mass of the planet.} To investigate the linearity of this relation, the Pearson correlation coefficient $\rm Pe_{c}$ between the mass of the planet $\rm M_{p}$ and the age of the system, for a given $\left\{ \rm P_{rot,i};\rm P_{orb,i};\rm M_{\star} \right\}$ triplet, is extracted from each synthetic pre-compiled age exploration file.
The Pearson correlation coefficient is a quantity that estimates the linear correlation between two variables; the closer this value is to one, the higher the linear correlation between these two variables is. Figure \ref{density} shows that most of the Pearson correlation coefficients{, extracted from the whole pre-compiled exploration grid,} are above 0.5 and close to 1. More than 64\% of the Pearson coefficient estimates calculated from the exploration files are above the moderate positive limit of 0.5, and the median of these estimates is 0.6 (average = 0.58). This suggests an actual linear correlation between $\rm M_{p}$ and the age of the system. 

To fully assess the validity of the tidal-chronology technique, age estimates provided by TATOO should be compared to fundamentally independent age predictions such as ages based on asteroseismic data \citep[see][and all the work of Dr Bellinger related to asteroseismic age]{Bellinger19b}. {Unfortunately, this test will have to wait for the TESS (Transiting Exoplanet Survey Satellite) and PLATO (PLAnetary Transits and Oscillations of Stars). After that please use only the abbreviation. An abbreviation or acronym may be introduced if it is used more than three times in the Abstract and more than five times in the main text (otherwise spell out).} missions, since for now only a few asteroseismic age estimations exist the host stars of massive close-in planets. Indeed, most of these estimates are for red giant branch stars or for Kepler targets that are composed of either a planet that is not sufficiently massive  or one whose semi-major axis is too large.} {Moreover, most of the targets listed in Table \ref{tabsystems} are located in the southern hemisphere, while the Kepler observations and the asteroseismic data are for targets in the northern hemisphere. Hence, PLATO and the future TESS observations are fundamental to constrain and test the tidal-chronology technique.} {It is also worth noting that sometimes only the minimum mass of the planet can be estimated, this is the case for instance in a radial velocity survey. In those cases, only a lower age limit could be estimated via TATOO.}

Finally, it is important to be aware that even if the principle of tidal-chronology is a physical reality \citep[see][]{Gallet19}, the outcome of this technique is only valid in the framework of the considered hypothesis regarding the geometry of the system (coplanar and circular orbit) as well as in relation to how the tidal interaction is treated. Moreover, because of the nature of the initial grid delivered with this version, TATOO currently only works for planetary systems composed of a star between 0.5 and 1.0 $\rm M_{\odot}$ around which orbits a planet between 0.5 and 3.5 $\rm M_{jup}$.
\begin{figure*}[]
        \centering
        \includegraphics[width=0.85\linewidth]{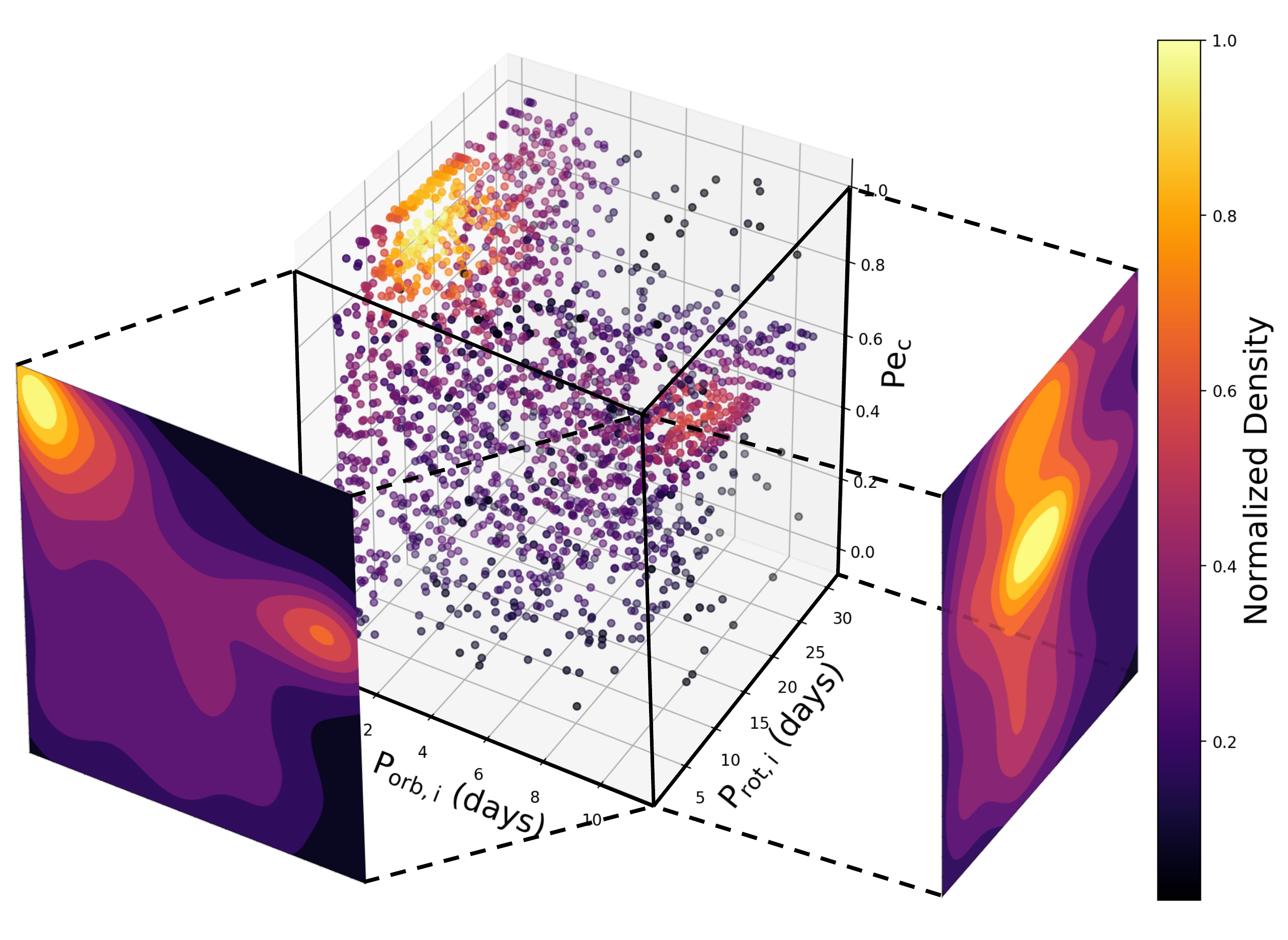}
    \caption{Three-dimensional plot between $\rm P_{orb,i}$, $\rm P_{rot,i}$, and the Pearson coefficient $\rm Pe_{c}$. The two additional two-dimensional plots display the density map of $\rm P_{orb,i}$ against $\rm Pe_{c}$ (left) and $\rm P_{rot,i}$ against $\rm Pe_{c}$ (right).}
    \label{density}
\end{figure*}

\subsection{Simple gyrochronological mode}

I first investigate whether TATOO can be used as a simple gyrochronological tool. For this, the ages of the observed systems listed in Table \ref{tabsystems} are extracted using the same procedure as above but by considering a higher $\rm P_{orb,obs}$ for the planet so as to remove the impact of this latter on the surface rotation of the star. {These ages, hereafter named tidal-gyrochronological ages, are thus the counterpart of the empirical gyrochronological ages but are estimated using the full \citet{Gallet18} period and orbital evolution numerical model.} These tidal-gyrochronological ages ($\rm Age_{tidal-gyro}$) are listed in Table \ref{tabsystems_gyro}. For the gyrochronological ages, the tool and calibration provided by \citet{Angus15} were used. They are calibrated using the Kepler asteroseismic targets
\begin{eqnarray}
\rm Age_{gyro} = \left( \frac{P_{rot,obs}} { 0.4(\rm B-V-0.45)^{0.31} }  \right)^{1/0.55}~Myr,
\end{eqnarray}
with the (B-V)-mass relation that is given by the YREC (The Yale Rotating Stellar Evolution Code) isochrones extracted at 600 Myr for solar metallicity stars \citep[Z = 0.01757, see][]{YREC}\footnote{http://www.astronomy.ohio-state.edu/iso/empirical.html}. {It is worth noticing that gyrochronology analysis can only be applied to systems with an actual age older than about 100-200 Myr for a 1.0 $\rm M_{\odot}$ star and 500-600 Myr for a 0.6 $\rm M_{\odot}$ star \citep{Barnes10,Delorme11}.} {The gyrochronological age estimates from \citet{Angus15} are on average 10\% smaller than the gyrochronological ages from \citet{Delorme11} used in \citet{Gallet19}.} {Each gyrochronological age estimate is the median of 100 evaluations using a random draw in the range of the observed $\rm \sigma_{P_{rot,\star}}$ of the actual stellar rotation period as the input rotation period; the error is given by the standard deviation of these 100 age estimations.}

{In Table \ref{tabsystems_gyro}, the {absolute} deviation between $\rm Age_{gyro}$ and $\rm Age_{tidal-gyro}$ is on average 15\%.  
From these estimates, 34\% have an {absolute} deviation smaller than 10\%, 53\% have an {absolute} deviation smaller than 15\%, 70\% have an {absolute} deviation smaller than 20\%, and 85\% have an {absolute} deviation smaller than 25\%.} {The agreement between $\rm Age_{gyro}$ and $\rm Age_{tidal-gyro}$ is better by using the \citet{Delorme11} calibration. In that case, 70\% of the sample have an {absolute} deviation lower than 10\%, and 92\% lower than 25\%.} Table \ref{tabsystems_gyro} shows that there is a good agreement between the two techniques, gyrochronology and tidal-gyrochronology, which is interesting since these two methods are fundamentally opposed. Indeed, gyrochronology is based on observations (empirical method) while tidal-chronology is strongly based on a theoretical approach. 

{The apparent agreement between these two age estimate paradigms validates the theoretical age determination employed by TATOO and shows that the model correctly works.} {It is even more interesting as empirical relations, such as gyrochronology, should in the long term be replaced by a theoretical-based model in which the physics is understood.}

\subsection{Application to several known systems}
\label{systems}

In this section, TATOO is used to estimate the age of several massive close-in planetary systems. These systems mostly come from \citet{Maxted15}, in which they compared isochrones to gyrochronological ages for various star-planets systems, and from \citet{Penev18}. In \citet{Maxted15}, they already highlighted the possible impact of tidal spin-up induced by the presence of a massive planet to explain the large discrepancies between isochrone and gyrochronological ages.

The properties of the planetary systems considered in this article are displayed in Table \ref{tabsystems}. This table also lists their estimated tidal-chronology and gyrochronology ages. {Systems for which the tidal-age is greater than the gyrochronological age by 20\% are considered to have been impacted by tidal spin-up.} {It seems that {a bit less than} half of these systems have a tidal age estimate close to the gyrochronological one.
The quality factor of each age estimation is defined as the sum between a value that depends on the value of $\rm Pe_{c}$ and the ratio between the age estimate and its standard deviation $\rm \sigma_{Age}$:
\begin{eqnarray}
\rm Quality =
\frac{Age}{\sigma_{Age}} + \left\lbrace
\begin{array}{ccc}
30 & \mbox{\rm if} & \rm Pe_{c} <0.5\\
60 & \mbox{\rm if} & 0.5<\rm Pe_{c}<0.8\\
90 & \mbox{\rm if} & \rm Pe_{c} \geq 0.8
\end{array}\right. .
\end{eqnarray}
The greatness of the age estimation increases with the value of the quality factor. {A quality factor above 60 indicates an intermediate quality while a quality factor of 90 is already a good estimate.}
\begin{table}
        \caption{Median precision of TATOO given the initial input conditions.}
        \label{precision}
        \centering
        \begin{tabular}{|c|c|c|}
        \hline 
        $\rm M_{\star}$ $\backslash$ $\rm M_{p}$  & 1 $\rm M_{jup}$ & 3 $\rm M_{jup}$ \\ 
        \hline 
        \multicolumn{3}{c|}{$\displaystyle \sigma_{\rm P_{orb,obs}}$ = $\displaystyle \sigma_{\rm P_{rot,obs}}$ = 1, 3, 5 and 7\%} \\
        \hline
        0.5 $\rm M_{\odot}$ & 1.9 / 5.1 / -- / -- & 2.5 / 7.9 / 43.2 / 53.2 \\ 
        \hline  
        0.6 $\rm M_{\odot}$ & 1.8 / 12.5 / 68.7 / 79.8 & 3.8 / 13.1 / 57.7 / 69.7 \\ 
        \hline 
        0.7 $\rm M_{\odot}$ & 1.4 / 4.8 / 18.7 / 23.9 & 2.3 / 23.16 / 66.2 / 73.4 \\ 
        \hline  
        0.8 $\rm M_{\odot}$ & 1.2 / 9.4 / 20.5 / 29.5 & 2.4 / 15.0 / 46.8 / 75.0 \\ 
        \hline  
        0.9 $\rm M_{\odot}$ & 1.3 / 9.5 / 20.5 / 26.6 & 1.3 / 5.8 / 15.7 /23.9 \\ 
        \hline
        1.0 $\rm M_{\odot}$ & 1.27 / 7.2 / 15.5 / 59.5 & 2.1 / 7.1 / 16.8 / 25.5 \\ 
        \hline   
        \end{tabular} 
        \begin{tablenotes}
      \small
      \item \textbf{Notes.} The values are extracted for $\displaystyle \sigma_{\rm P_{orb,obs}}$ = $\displaystyle \sigma_{\rm P_{rot,obs}}$ = 1, 3, 5, and 7\%, respectively.

    \end{tablenotes}    
\end{table}
{Table \ref{precision} shows the values of the precision of the age estimates extracted from TATOO as a function of the stellar and planetary mass, and on the input errors $\displaystyle \sigma_{\rm P_{rot,obs}}$ and $\displaystyle \sigma_{\rm P_{orb,obs}}$. With an error below 3\% on $\rm P_{rot,obs}$ and $\rm P_{orb,obs}$ the precision of TATOO is lower than 10\%. For a given value of $\displaystyle \sigma_{\rm P_{orb,obs}}$ and $\displaystyle \sigma_{\rm P_{rot,obs}}$, the precision of TATOO increases for increasing rotation period and orbital period.}

\section{Discussion}
\label{discussion}

The apparent existence of the linear relation between the age and the planetary mass of the system suggests that for some systems the heavier the planetary mass is, the older the system. It means that for two given planetary systems that are subject to tidal interaction, if they have the same set of observed properties then their age is proportional to the mass of their orbiting planet. Indeed, since  the orbital evolution timescale $\Ts$ evolves as $\rm a^8$ \citep[see][]{Bolmont16}, then for a given $\rm P_{orb,obs}$ and stellar properties, if two massive planets are still observed orbiting close to their central star, then it means that the more massive planet has started its evolution at a greater $\rm P_{orb,init}$ than the less massive one, and that therefore this system is the oldest one. 
\begin{figure}[]
        \centering
        \includegraphics[width=\linewidth]{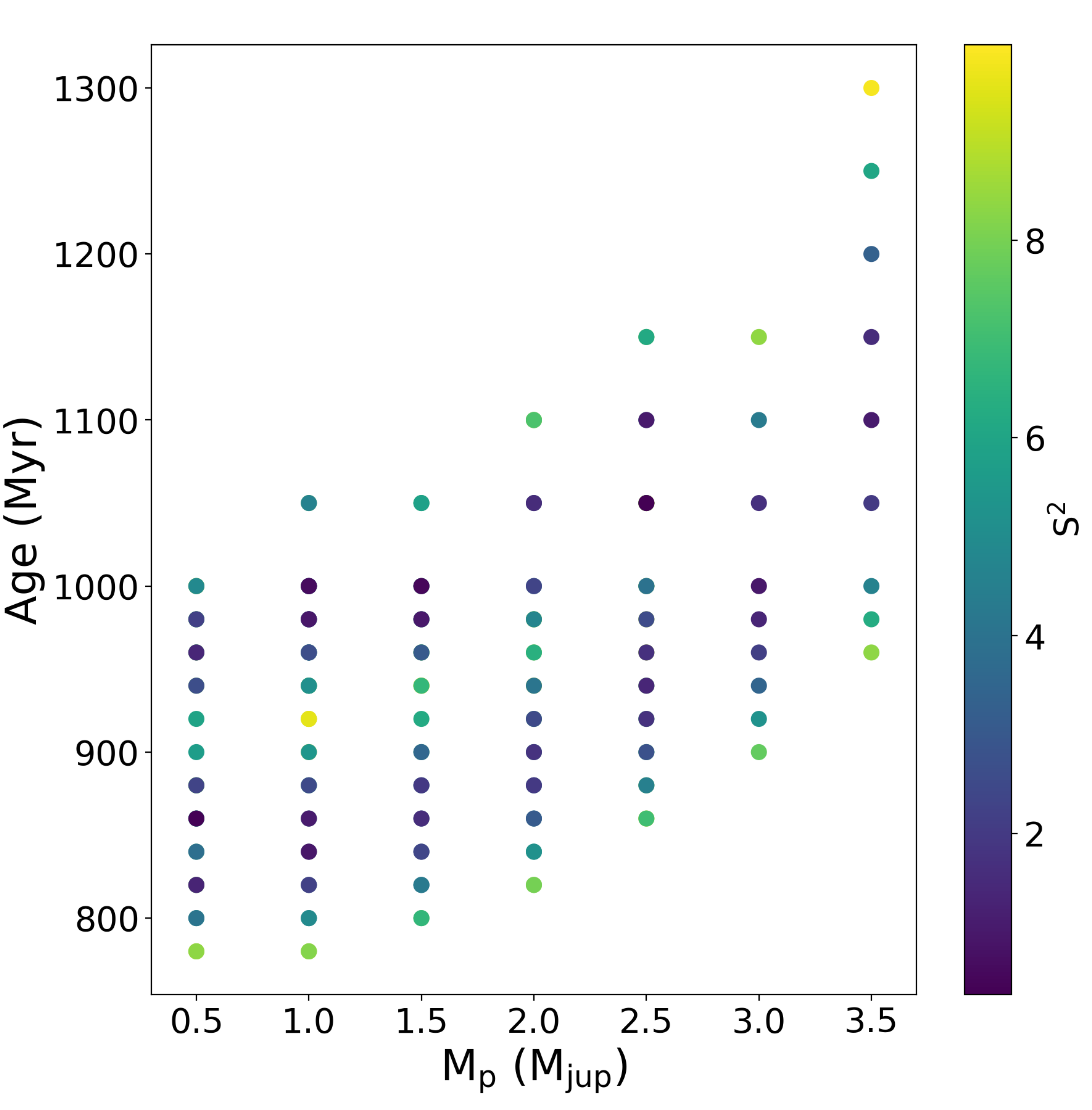}
    \caption{Estimated age of a system composed of a 0.7 $\rm M_{\odot}$ with a rotation of 16 days as a function of the mass of the planet that is located at 0.02 au.}
    \label{relation}
\end{figure}
{For example, say that we want to get the age a system in which the star is a 0.7 $\rm M_{\odot}$ with a rotation of 16 days and that there is a planet of unknown mass orbiting this latter at 0.02 au. At such a small star-planet distance, we expect that the planet will be impacted by tidal interaction that will most probably induce inward migration. Since more massive planets evolve faster because the characteristic timescale $\rm T_{\star}$ decreases for increasing planetary mass, if two systems have the same properties  (same stellar mass $\rm M_{\star}$, same stellar rotation period $\rm P_{rot,\star}$, and same planetary orbital period $\rm P_{orb,p}$) then the age of the system increases for increasing planetary mass. This effect is depicted in Fig. \ref{relation}, which shows the estimated age of a system composed of a 0.7 $\rm M_{\odot}$ with a rotation period of 16 days as a function of the mass of the planet that is located at 0.02 au around the star. The colour gradient depicts the value of the $\rm S^2$.}

{In the sample explored in this article, half of the systems experienced tidal-spin up during their evolution. These systems are the ones, on average, with the highest ratio $\rm P_{rot,\star}$/$\rm P_{orb,p}$ and $\rm R_{\rm co}$/$a$, where $\rm R_{\rm co}$ is the corotation radius defined as}
\begin{equation}
\label{corot}
\rm R_{\rm co} = \left( \displaystyle \frac{\G M_{\star} P_{rot,\star}^2 }{(2\pi)^2}  \right)^{1/3},
\end{equation}
{which is the distance at which the orbital period of the planet equals the rotation period of the star}. {For the tidal spin-up cases the averaged $\rm P_{rot,\star}$/$\rm P_{orb,p}$ and $\rm R_{\rm co}$/$a$ ratios are around 9.2 and 1.24, respectively. For the non tidal spin-up cases the averaged $\rm P_{rot,\star}$/$\rm P_{orb,p}$ and $\rm R_{\rm co}$/$a$ ratios are around 6.0 and 0.98, respectively.} {A ratio  of $\rm R_{\rm co}$/$a$ > 1.0 refers to a planet inside the corotation radius, which may indicate that this latter has had an impact on the surface rotation rate of its host's star. In all of the explored cases, $\rm P_{orb,p}$ is smaller than $\rm P_{rot,\star}$.} 
\begin{figure}[]
        \centering
         \includegraphics[width=\linewidth]{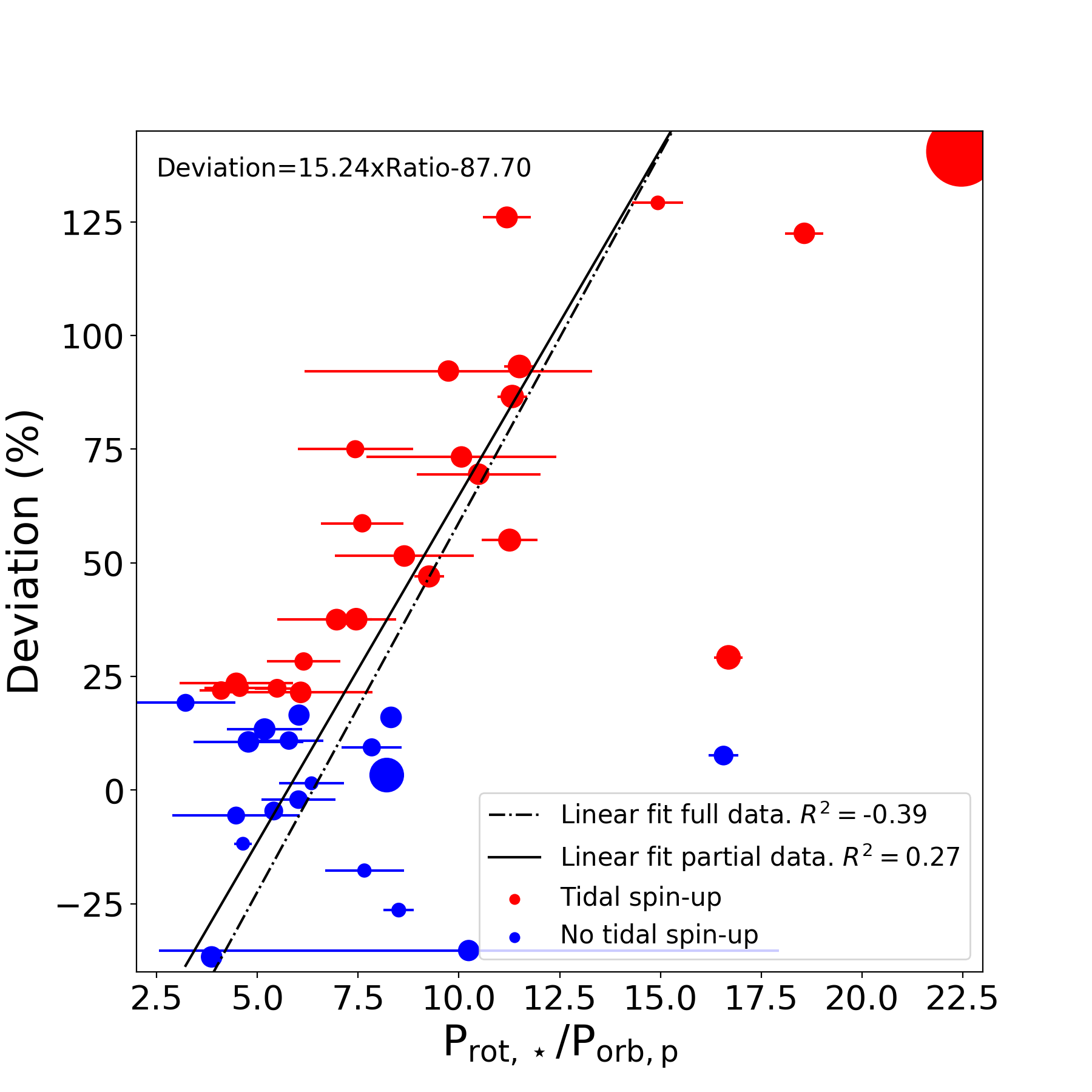}
    \caption{Deviation between tidal-chronology age and gyrochronology age as a function of the ratio $\rm P_{rot,\star}$/$\rm P_{orb,p}$. The dashed line is a linear fit considering the full dataset. The dash-dotted line is also a linear fit but omitting the WASP-4 and Qatar-1 systems from the dataset. The red dots are the systems for which tidal-spin up is suspected while the blue dots are systems without tidal spin-up. The dashed line corresponds to a linear fit to the full data while the dash-dotted line is a linear fit to the same data but without the WASP-4 and Qatar-1 systems that are located in the lower-right corner of the plot (with a ratio \textasciitilde\ 16\% and a deviation lower about 7 and 30\%, respectively). The size of the points increases for increasing quality factor extracted from Table \ref{tabsystems}.}
    \label{ratio}
\end{figure}
Figure \ref{ratio} shows the deviation between the age estimated with tidal-chronology and the one using gyrochronology as a function of the ratio $\rm P_{rot,\star}$/$\rm P_{orb,p}$. The quantity $R^2$ is the coefficient of determination defined by
\begin{equation}
\displaystyle R^2 = 1 - \frac{\sum\limits_{i=1}^n \left( y_i - f_i \right)^2}  {\sum\limits_{i=1}^n \left( y_i - \bar{y} \right)^2},
\end{equation}
{with $y_i$ the data, $f_i$ the linear fit to the data, and $\bar{y}$ the data's averaged value. A perfect linear fit gives $R^2= 1$.} This figure shows that the deviation, which is linked to tidal spin-up, scales linearly with the ratio $\rm P_{rot,\star}$/$\rm P_{orb,p}$: the higher this latter is, the stronger the deviation will be. The mass of the systems, both stellar and planetary, does not seems to affect this linearity. {This behaviour can be explained by the fact that an increasing $\rm P_{rot,\star}$/$\rm P_{orb,p}$ corresponds to an increasing $\rm R_{\rm co}$/$a$ quantity that is linked to systems in which the planet is located increasingly close inside the corotation radius. Therefore, a high $\rm P_{rot,\star}$/$\rm P_{orb,p}$ ratio system is expected to have experienced stronger tidal interaction and thus more powerful tidal spin-up effect.} From the linear fit in Fig. \ref{ratio} we can infer a limit around $\rm P_{rot,\star}$/$\rm P_{orb,p}$ = 7 (corresponding to $a$ \textasciitilde\ 1.01 $\rm R_{\rm co}$) above which the deviation is larger than 20\% and therefore tidal-chronology should be used. {This trend could then suggest that either the orbital and rotational period of the WASP-4 and Qatar-1 systems are incorrect or that the deviation and age provided by TATOO for these systems are well underestimated. If this latter hypothesis is true, then these systems should be placed in the tidal spin-up region for which gyrochronological ages are biased.} 
In the case of WASP-4, it seems that this system is highly sensitive to the $\rm S^2_{lim}$ value. Indeed, Table \ref{s2_lim} shows that the deviation for this system could reaches 50\% with $\rm S^2_{lim} =18$. However, the Qatar-1 system appears to be stable regarding the choice of $\rm S^2_{lim}$ , which could point towards a incorrectly estimated rotational period (assuming that the orbital period is correct). \citet{Covino13} reported a rotational period of 24$\pm$6 days for Qatar-1, which is very close to the 23.7$\pm$0.5 days found by \citet{Penev18} but with a higher deviation. This produces a tidal-age of 3728 Myr corresponding to a deviation of about 40\%. Given the orbital period of Qatar-1 b of 1.42 days, the ratio $\rm P_{rot,\star}$/$\rm P_{orb,p}$ could range from 12.7 to 21, placing the Qatar-1 system in better agreement with the linear model.

{Finally TATOO, like gyrochronology, is also not well suited for measuring the age of binary systems in which an external torque due to the stellar companion could have been applied to the surface rotation of the primary body. This is probably the case in this work for the TrES-2 system, which is known to be a binary system. For this system the tidal-gyrochronological age differs by -35.6\% from the simple gyrochronological estimation. For this kind of system, the estimated age using tidal-chronology is most probably biased depending on the strength of the binary interaction.} {Removing this system from the sample increases the linearity of the fit up to $\rm R^2 = 0.5$ and reduces the tidal spin-up limit down to 6.74 corresponding to $a$ \textasciitilde\ 0.98 $\rm R_{\rm co}$.}

\section{Conclusion}
\label{conclusion}

While the age of stars is a fundamental parameter that can strongly constrain on-going stellar and planetary models, it is currently the most uncertain one. It is especially true in the case of stars that are or were subject to tidal spin-up induced by the presence of a massive close-in companion. In those specific cases, gyrochronology analysis cannot be used without having a bias towards a younger age estimate. This apparent young age will then lead to an incorrect estimation of the system properties.

To solve this problem, this paper provides the astrophysical community with a standalone tool, based on the initial work of \citet{Gallet19}, which can be used to rapidly and easily get an estimate of the age of a given massive close-in planetary system. This method only requires information on the stellar and planetary fundamental parameters: $\rm M_{\star}$, $\rm M_{p}$, $\rm P_{rot,obs}$, $\rm P_{orb,obs}$. This interconnectivity between stellar and planetary characteristics clearly highlights the need for the stellar and planetary communities to systematically provide all the available properties of the star and planet of their systems of interest. 

While the most recent techniques only consider isolated stars \citep{Angus19,Bellinger19}, TATOO is currently the only tool that proposes to estimate the age of massive close-in planetary systems by taking into account the possible impact of the star-planet tidal interaction on the rotational evolution of the stellar surface. {This tool can also be used even if no rotational departure is present. In that case, it gives results in agreement with the classical gyrochronological analysis.} However, as with isochrone fitting, TATOO is model dependent and relies on the physics adopted in the numerical model used to produce the initial grid \citep[here PROBE from][]{Gallet18}. Hence, this model dependence should be taken into account when choosing to use TATOO as an age estimation tool. In this work I found that half of the investigated massive close-in systems for which TATOO can be applied, are subject to tidal spin-up and therefore have biased gyrochronological age estimations. I point out that this bias strongly scales with the ratio $\rm P_{rot,\star}$/$\rm P_{orb,p}$ , which therefore could be used to easily predict the correctness of the gyrochronological age estimates. According to the investigated systems, a ratio of $\rm P_{rot,\star}$/$\rm P_{orb,p} \gtrsim$ 7 is the indication that tidal-chronology should be preferred to a simple gyrochronological tool. Indeed, above this limit the deviation between the two techniques starts to be greater than 20\%. {A list of such systems is given in Table \ref{system_tidal}}. Moreover, correct ages can be retrieved, even without tidal spin-up effect, by using TATOO. In that case TATOO provides age estimates in good agreement with the gyrochronology technique. This proves that tidal-chronology can be seen as a first-order correction of the impact of tidal interaction on gyrochronology. 

\begin{table}
\caption{List of systems that may benefit from tidal-chronology over simple gyrochronology analysis.}
\label{system_tidal}
\centering
\begin{tabular}{|c|c|}
        \hline 
        Name & $\rm P_{rot,\star}$/$\rm P_{orb,p}$ \\
\hline 
\hline
WASP-80 & 7.65$\pm$0.98 \\ 
Qatar-2 & 8.51$\pm$0.37 \\ 
TrES-5 & 7.84$\pm$0.74 \\ 
WASP-4 & 16.57$\pm$0.37 \\ 
TrES-2 & 10.24$\pm$7.69 \\ 
WASP-50 & 8.32$\pm$0.26 \\ 
Kepler-423 & 8.21$\pm$0.04 \\ 
WASP-19 & 14.94$\pm$0.63 \\ 
WASP-135 & 7.43$\pm$1.43 \\ 
HAT-P-53 & 7.60$\pm$1.02 \\ 
WASP-36 & 9.74$\pm$3.57 \\ 
WASP-64 & 10.06$\pm$2.36 \\ 
WASP-43 & 18.57$\pm$0.48 \\ 
WASP-5 & 10.49$\pm$1.53 \\ 
HATS-9 & 8.65$\pm$1.72 \\ 
HAT-P-43 & 6.97$\pm$1.47 \\ 
HATS-18 & 11.19$\pm$0.60 \\ 
HATS-2 & 9.26$\pm$0.37 \\ 
HATS-33 & 7.45$\pm$0.39 \\ 
WASP-46 & 11.26$\pm$0.70 \\ 
WASP-77-A & 11.32$\pm$0.37 \\ 
HAT-P-36 & 11.50$\pm$0.38 \\ 
Qatar-1 & 16.69$\pm$0.35 \\ 
NGTS-10 & 22.47$\pm$0.52 \\
\hline
\end{tabular} 
\end{table}
Finally, in the framework of the next generation of space missions such as PLATO \citep{Plato} and TESS \citep{TESS}, TATOO will be a valuable age estimation tool as it will be crucial to have the best information on the planetary systems chosen as targets, including the estimation of their age, to fully interpret the data. {The recent work of \citet{Amard20} also shows that metallicity could have a non-negligible effect for example on the validity of the current gyrochronology calibrations. This suggests the need for a more detailed analysis of the stability of tidal-chronology regarding the metallicity of the planetary systems investigated. This should be done in future works and is currently beyond the aim of this present one.}

\begin{acknowledgements}
{F.G. thanks the anonymous referee for the constructive comments about this work that increased the quality of the paper.} F.G. thanks Dr. Marta Gonz\'{a}lez Garc\'{i}a for her precious help with statistical analysis and her comments on the article. F.G. is grateful to Dr. Philippe Delorme, Dr. Jacques Kluska, and Dr. Julien Rameau for helpful discussions and corrections regarding the content of the article. F.G acknowledges financial support from the CNES fellowship. This project has received funding from the European Research Council (ERC) under the European Union's Horizon 2020 research and innovation programme (grant agreement No 742095; {\it SPIDI}: Star-Planets-Inner Disk-Interactions). 
\end{acknowledgements}

\bibliographystyle{aa} 
\bibliography{references} 

\appendix

\section{Additional tables and figures}

Table \ref{s2_lim} shows the impact on the age estimation of the choice of $\rm S_{lim}^2$. {It shows that the difference can reach up to 79\%, but 90\% of the estimated ages have a deviation lower or equal to 10\%.}

\begin{table}
        \caption{Impact of the choice of $\rm S_{lim}^2$ on the age estimation with TATOO.}
        \label{s2_lim}
        \centering
        \begin{tabular}{|c|c|c|c|}
        \hline 
        Name &   $\rm Age_{tidal,S_{lim}^2=9}$ &  $\rm Age_{tidal,S_{lim}^2=18}$  &  Deviation \\ 
                                &       (Myr) & (Myr) & \% \\ 
\hline
WASP-43 & 3198.67$\pm$1460.76 & 1541.18$\pm$64.4 & -51.82 \\ 
WASP-19 & 3789.05$\pm$403.96 & 2631.35$\pm$24.17 & -30.55 \\ 
HATS-18 & 3392.57$\pm$94 & 2789.75$\pm$118.69 & -17.77 \\ 
HATS-9 & 3544.13$\pm$1068.59 & 3159.83$\pm$1060.51 & -10.84 \\ 
WASP-65 & 1563.11$\pm$285.4 & 1442.87$\pm$293.8 & -7.69 \\ 
WASP-36 & 5710.38$\pm$4125.43 & 5289.89$\pm$3674.27 & -7.36 \\ 
HATS-15 & 709.45$\pm$141.04 & 657.59$\pm$175.37 & -7.31 \\ 
WASP-64 & 3554$\pm$1628.82 & 3297.22$\pm$1205.19 & -7.23 \\ 
HATS-14 & 1295.62$\pm$597.71 & 1205.6$\pm$587.14 & -6.95 \\ 
WASP-44 & 1691.76$\pm$468.93 & 1588.97$\pm$512.91 & -6.08 \\ 
CoRoT-13 & 2013.6$\pm$1008.75 & 1911.95$\pm$1031.98 & -5.05 \\ 
HATS-34 & 1065.66$\pm$167.22 & 1016.07$\pm$154.55 & -4.65 \\ 
HAT-P-36 & 4888.13$\pm$212.75 & 4675.54$\pm$153.09 & -4.35 \\ 
WASP-46 & 2152.68$\pm$127.47 & 2059.56$\pm$131.43 & -4.33 \\ 
NGTS-10 & 6116.81$\pm$5.57 & 5857.33$\pm$454.6 & -4.24 \\ 
HAT-P-53 & 3886$\pm$689.09 & 3734.38$\pm$724.17 & -3.9 \\ 
HATS-30 & 2027.22$\pm$292.97 & 1963.95$\pm$272.81 & -3.12 \\ 
HAT-P-37 & 1484.68$\pm$301.29 & 1451.41$\pm$292.64 & -2.24 \\ 
WASP-77-A & 4309.1$\pm$203.53 & 4221.16$\pm$135.25 & -2.04 \\ 
HATS-2 & 1357.39$\pm$34.52 & 1335.08$\pm$18.79 & -1.64 \\ 
HD189733 & 663.34$\pm$17.17 & 657.74$\pm$0.79 & -0.84 \\ 
CoRoT-29 & 1453.45$\pm$292.49 & 1441.68$\pm$316.79 & -0.81 \\ 
Qatar-2 & 412.73$\pm$40.78 & 410.19$\pm$38.5 & -0.62 \\ 
WASP-10 & 419.43$\pm$12.78 & 418.2$\pm$8.56 & -0.29 \\ 
Kepler-423 & 2317.8$\pm$13.6 & 2318.11$\pm$14.47 & 0.01 \\ 
WASP-41 & 2289.4$\pm$77.39 & 2294.88$\pm$74.23 & 0.24 \\ 
WASP-57 & 956.65$\pm$518.41 & 961.09$\pm$566.52 & 0.46 \\ 
WASP-23 & 1237.88$\pm$178.98 & 1247.93$\pm$196.57 & 0.81 \\ 
WASP-85 & 2063.06$\pm$231.36 & 2082.51$\pm$232.79 & 0.94 \\ 
WASP-50 & 1662.38$\pm$50.35 & 1681.01$\pm$62.98 & 1.12 \\ 
HATS-33 & 4260.12$\pm$328.49 & 4320.03$\pm$335.62 & 1.41 \\ 
WASP-80 & 1341.26$\pm$187.27 & 1364.67$\pm$191.15 & 1.75 \\ 
HAT-P-43 & 5966.69$\pm$1345.24 & 6113.35$\pm$1285.77 & 2.46 \\ 
WASP-140 & 563.9$\pm$123.9 & 580.15$\pm$119.94 & 2.88 \\ 
Qatar-1 & 3376.92$\pm$36.24 & 3590.78$\pm$188.79 & 6.33 \\ 
WASP-135 & 1811.18$\pm$553.8 & 1948.09$\pm$533.4 & 7.56 \\ 
WASP-5 & 3725.97$\pm$1264.42 & 4057.6$\pm$940.63 & 8.9 \\ 
TrES-5 & 872.18$\pm$247.42 & 960.63$\pm$260.96 & 10.14 \\ 
WASP-124 & 2642.69$\pm$1077.12 & 2983.05$\pm$975.34 & 12.88 \\ 
TrES-2 & 2353.87$\pm$2698.34 & 3512.1$\pm$2975.47 & 49.21 \\ 
WASP-4 & 2694.94$\pm$164.18 & 4828.06$\pm$52.16 & 79.15 \\
        \hline   
        \end{tabular} 
\end{table}

Figure \ref{initialgrid} is a schematic description of the composition of the initial grid.

\begin{figure}[!ht]
    \begin{center}
            \includegraphics[width=\linewidth]{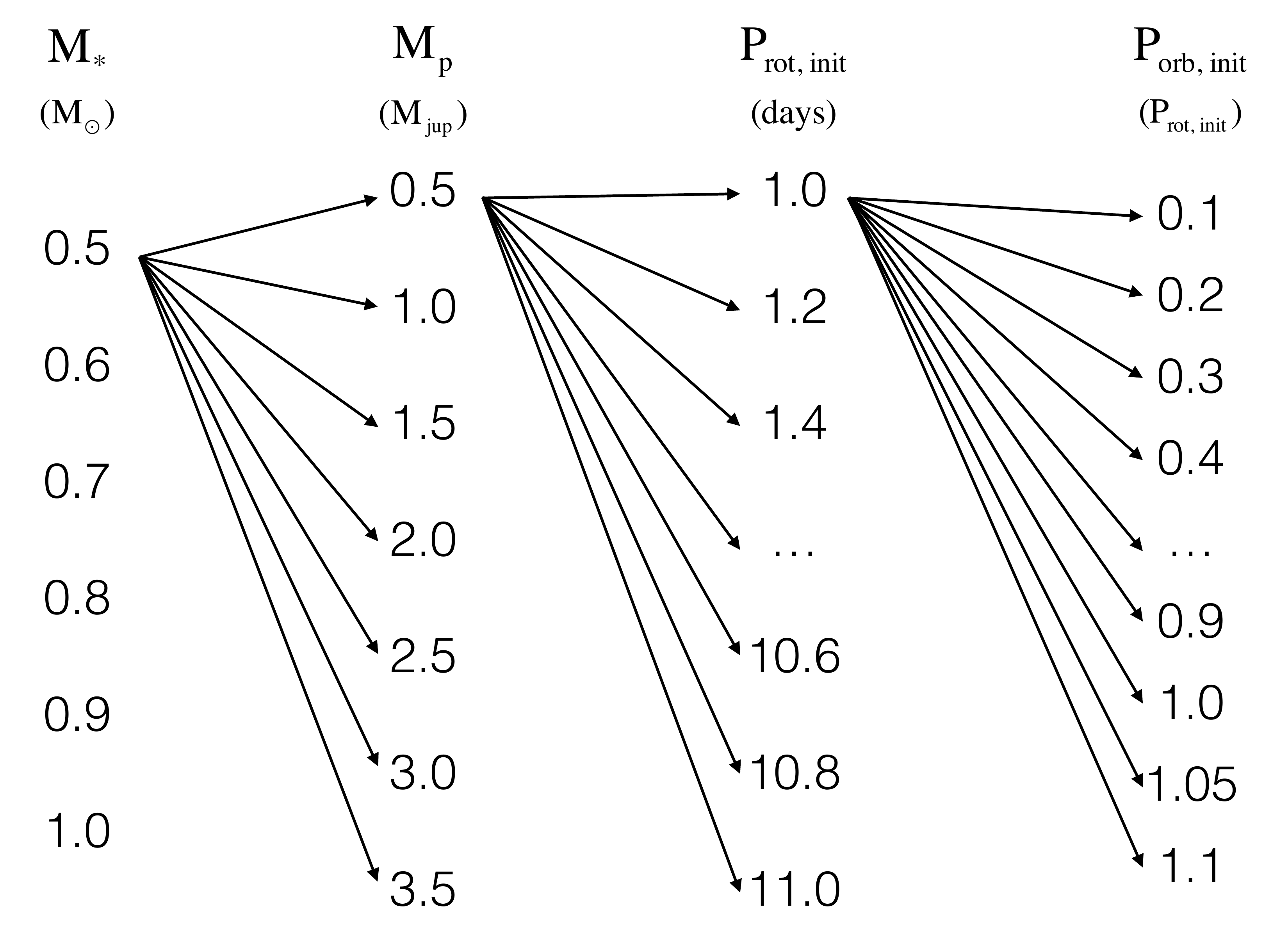}
        \caption{Schematic description of the initial grid.}
        \label{initialgrid}%
    \end{center}
\end{figure}

\end{document}